\documentclass[12pt]{article}

\usepackage{amsfonts,amsmath,amssymb,amsthm}
\usepackage[hyperfootnotes=false,bookmarks=false]{hyperref}

\font\tenmath=msbm10 scaled 1200
\font\sevenmath=msbm7 scaled 1200
\font\Fivemath=msbm5 scaled 1200
\newfam\mathfam \textfont\mathfam=\tenmath
\scriptfont\mathfam=\sevenmath \scriptscriptfont\mathfam=\Fivemath

\newtheorem{theorem}{Theorem} 
\def \\ { \cr }
\def\R{\mathbb{R}}
\def \1{1 \mkern -6mu 1} 
\def\N{\mathbb{N}}
\def\P{\mathbb{P}}

\def\Z{\mathbb{Z}}
\def\R{\mathbb{R}}
\def \e{{\rm e}}

\begin{document}

\title{Elephant Random Walks and their connection to P\'olya-type urns}
\author{{Erich Baur\footnote{erich.baur@ens-lyon.fr} { and } Jean
    Bertoin\footnote{jean.bertoin@math.uzh.ch}}\\ ENS Lyon and 
  Universit\"at Z\"urich}
 \date{}
\maketitle 
\thispagestyle{empty}
\begin{abstract}
  In this paper, we explain the connection between the Elephant Random Walk
  (ERW) and an urn model \`a la P\'olya and derive functional limit
  theorems for the former. The ERW model was introduced by Sch\"utz and
  Trimper~\cite{ScTr} in 2004 to study memory effects in a one-dimensional
  discrete-time random walk with a complete memory of its past. The
  influence of the memory is measured in terms of a parameter $p$
  between zero and one. In the past years, a considerable effort has been
  undertaken to understand the large-scale behavior of the ERW,
  depending on the choice of $p$. Here, we use known results on urns to
  explicitly solve the ERW in all memory regimes. The method works as well 
  for ERWs in higher dimensions and is widely applicable to related
  models. 
\end{abstract}

\footnote{{\it Acknowledgment of support.} The research of EB
  was supported by the Swiss National Science Foundation grant
  P300P2\_161011, and performed within the framework of the LABEX MILYON
  (ANR-10-LABX-0070) of Universit\'e de Lyon, within the program
  ``Investissements d'Avenir'' (ANR-11-IDEX-0007) operated by the French
  National Research Agency (ANR).}

\section{Introduction}
Random walks and, more generally, diffusion processes are widely used in
theoretical physics to describe phenomena of traveling motion and mass
transport. Due to the fractal structure of nature and space and temporal
long-range correlations in particle movements (see,
e.g.,~\cite{Ma,MaNe,MeKl,ScMo,Wa}), so-called anomalous diffusions often 
appear, where the mean square displacement of a particle is no longer a
linear function of time, but is rather given by a power law.

A simple model exhibiting anomalous diffusion is the so-called Elephant
Random Walk (ERW) introduced by Sch\"utz and Trimper~\cite{ScTr} in 2004,
which is the topic of this paper. The ERW model is a one-dimensional
discrete-time nearest-neighbor random walk on $\Z$, which remembers its
full history and chooses its next step as follows: First, it selects
randomly a step from the past, and then, with probability $p\in[0,1]$, it
repeats what it did at the remembered time, whereas with the complementary
probability $1-p$, it makes a step in the opposite direction. We refer to
the next section for the precise definition. The memory parameter
$p\in[0,1]$ allows us to model the willingness of the walker to do the same as
in the past. When $p=1/2$, the memory has no effect on the movement: The
model coincides with the simple symmetric random walk.

The ERW model and some variations thereof have drawn a lot of attention in
the last years, see,
e.g.,~\cite{AlArCrSiSiVi,BoRC,HaKuLi,Ha1,Ha2,KuHaLi,Kue,PaEs,ScTr,Se,SiCrScViTr}
to mention just a few. One of the key questions concerns the influence of
the memory on the long-time behavior. Various results and predictions have
been obtained, e.g., in~\cite{PaEs,ScTr,SiCrScViTr}. In this paper, we
explicitly determine the long-time behavior of the ERW model in all regimes
$p\in[0,1]$. We obtain central limit theorems for the full process of the
ERW, with a scaling depending on the choice of $p$. In the regime
$p\leq 3/4$, the limiting process turns out to be Gaussian (with explicit
parameters). In the superdiffusive case $p>3/4$, the limit is non-Gaussian,
as it was already predicted in~\cite{SiCrScViTr, PaEs}. We point out that our 
limit theorems are stronger than finite-dimensional convergence of
the ERW. In particular, they imply convergence of continuous functionals of
the walker.

Our method uses a connection to P\'olya-type urns that was already known
before in the literature, see, e.g., the works of Harris~\cite{Ha1,Ha2} and
also the survey of Pemantle~\cite{Pe} on related random processes with
reinforcement. Being robust and simple, the method is neither limited to
one-dimensional models nor to the specific ERW model, but rather
widely applicable to other random walks with memory. A bit more precisely,
given what is known from the theory of urns, we will see that the
asymptotic behavior of such models is essentially determined by the
spectral decomposition of the (replacement) matrix of the corresponding
urn.

Since the ERW is arguably the most natural and simplest model of a 
one-dimensional random walk with a complete memory, we concentrate in this
paper on the basic ERW and leave it mostly to the reader to adapt the method
to other walks with memory. However, we outline some possible extensions in
Section~\ref{sec:extensions}.

The rest of this paper is structured as follows. After having introduced the
exact ERW model in the following section, we describe in
Section~\ref{sec:urn} a particular discrete-time urn model containing balls
of two colors, where step by step a new ball is added. We then show in
Section~\ref{sec:results} how the known limit results on the composition of
the urn can be transferred into statements about the position of the ERW
when time goes to infinity. In Section~\ref{sec:extensions}, we discuss
various extensions, and, in the last part, we summarize our findings.

We finally mention that independently of us and at the same time as ours, a
work of Coletti, Gava and Sch\"utz~\cite{CoGaSc} appeared on the arXiv,
with related results on the ERW but using a different approach.

\section{The model}
Let us now introduce the exact model, in the way it was first defined
in~\cite{ScTr}. The ERW is a one-dimensional random walk $(S_n,n\in\N_0)$
on the integers 
starting, say, at zero at time zero, $S_0=0$. At time $n\geq 1$, the
position of the walk is given by
$$
S_n = S_{n-1}+\sigma_n,
$$
where $\sigma_n$, $n\in\N=\{1,2,\ldots\}$, are random variables taking
values in $\{\pm 1\}$, which are specified as follows. Firstly, $\sigma_1$
takes the value $1$ with some probability $q\in [0,1]$ and the value $-1$ with
probability $1-q$. Accordingly, the first step of the ERW goes to the right
(left) with probability $q$ $(1-q)$. At any later time $n\geq 2$, we choose
a number $n'$ uniformly at random among the previous times ${1,\ldots,n-1}$
and set
$$
\sigma_n= \left\{\begin{array}{l@{\quad\mbox{with probability  }}l}
      +\sigma_{n'} & p\\
      -\sigma_{n'} &1-p
\end{array}\right.,
$$
where $p\in [0,1]$ is a memory parameter which is inherent to the
model. Note that the case $p=1/2$ corresponds to simple symmetric random
walk: there is no memory effect. Moreover, we remark that
$S_n=\sigma_1+\ldots+\sigma_n$. We implicitly agree that the various random
choices made in this construction are independent from each other.

In~\cite{ScTr}, the question of how the memory of the history influences
the position of the walker at large times was investigated. In particular,
by writing $\langle \cdot\rangle$ for the expectation operator, it was shown
that the mean displacement of the ERW satisfies for $n\gg 1$,
\begin{equation}
\label{eq:mean}
\langle S_n\rangle \sim \frac{(2q-1)}{\Gamma(2p)}n^{2p-1},
\end{equation}
while for the second moment, it was proved that 
\begin{equation}
\label{eq:meansquare}
\langle S_n^2\rangle \sim \left\{\begin{array}{l@{\quad\mbox{for }}l}
    \frac{n}{3-4p} & 0\leq p<3/4\\
    n\ln n&p=3/4\\
    \frac{n^{4p-2}}{(4p-3)\Gamma(4p-2)}&3/4<p\leq 1
\end{array}.\right.
\end{equation}
The last display entails at $p=3/4$ a transition from a diffusive
$(0\leq p<3/4)$ to a superdiffusive ($3/4<p\leq 1$) regime, whereas at
$p=3/4$, the ERW behaves marginally superdiffusive. Using an approximation
by a Fokker-Planck equation, the random walk propagator of
the ERW model was reported in~\cite{ScTr} to be Gaussian in all regimes
(with a time dependent diffusion constant), an observation which was later
adapted in~\cite{SiCrScViTr} for the superdiffusive regime $p>3/4$, where
a more precise analysis showed that the random walk propagator is in fact
non-Gaussian. Here, the term {\it propagator} refers to the probability
density of the usual continuum limit. See also~\cite{PaEs} for a related
work confirming that the Fokker-Planck approximations do not yield adequate
results for the ERW model, at least not in the superdiffusive regime. The
statistics in the regime $1/2<p\leq 3/4$ were left open
in~\cite{SiCrScViTr}.

The main purpose of this paper is to affirm the observation
of~\cite{SiCrScViTr} in the superdiffusive regime and clarify the behavior
in the remaining regimes, by explicitly calculating the large-scale
behavior of the ERW model by using a connection to P\'olya-type urns, which we
explain next.

\section{The connection to P\'olya-type urns}
\label{sec:urn}
Imagine a discrete-time urn with balls of two colors; say, black and
red. The composition of the urn at time $n\in\N$ is given by a vector
$X_n=(X_n^1,X_n^2)$, where the first component $X_n^1$ counts the number of
black balls at time $n$, and the second component $X_n^2$ counts the number of red
balls. We restrict ourselves to starting compositions $X_1=\xi$ for some
(possibly random) vector $\xi=(\xi^1,\xi^2)$ taking values in
$\{(1,0),\,(0,1)\}$ almost surely. The urn now evolves according to the
following dynamics: At time $n=2,3,\ldots$, we draw a ball uniformly at
random, observe its color, put it back to the urn and add with probability
$p$ a ball of the same color, and with probability $1-p$ a ball of the
opposite color. Then we update $X_n$, so that $X_n$ describes the
composition of the urn after the $(n-1)$st drawing.

The connection to the ERW model is remarkably
simple: If $(S_n,n\in\N_0)$ is the ERW started
from $S_0=0$ such that $S_1=\xi^1-\xi^2$, then
\begin{equation}
\label{eq:ERW-urn}
(S_n,n\in\N) =_d (X_n^1-X_n^2,n\in\N),
\end{equation}
where $=_d$ refers to equality in law. In other words, the difference
between the number of black and red balls in the above urn evolves like an
ERW with first step equaling $\xi^1-\xi^2$.

The urn described above fits into a broader setting of so-called
generalized Friedman's or P\'olya urns; see~\cite{Be1,Be2,Fr,Sa} for first
results (with deterministic replacement rules). Athreya and
Karlin~\cite{AtKa} proved an embedding of urn schemes into continuous-time
multitype Markov branching processes, which includes the treatment of
generalized Friedman's urn processes with randomized replacement rules, as
in our case. These techniques were further developed by Janson
in~\cite{Ja}, which serves as the main reference for this paper. Many
results on urns can also be found in Mahmoud's book~\cite{Mah}, which is,
however, more combinatorial in nature.

Key quantities that govern the long-time behavior of the urn
process are the eigenvalues and eigenvectors of the so-called mean
replacement matrix. In our case, it is given by
\begin{equation}
\label{eq:A}
A=\begin{pmatrix}p&1-p\\1-p&p\end{pmatrix}.
\end{equation}
The eigenvalues of $A$ are $\lambda_1=1$, $\lambda_2=2p-1$, and
the corresponding right and left eigenvectors are $v_1=\frac{1}{2}(1, 1)'$,
$v_2=\frac{1}{2}(1, -1)'$, $u_1=(1,1)$, $u_2=(1,-1)$, where we write $v'$
for the transpose of $v$. Here, as in $(2.2)$ and $(2.3)$ of~\cite{Ja}, we
have chosen $v_1,v_2$ and $u_1,u_2$ such that $u_1v'_1 = u_2v'_2=1$ and the
$L^1$-norm of $v_1$, $v_2$ is equal to one. 

It is well-known (see, e.g.,~\cite{AtKa, ChPoSa,KeSt,Ja}) that the
asymptotics of the urn depends on the position of $\lambda_2/\lambda_1$
with respect to $1/2$ (in the situation of a more general urn, assuming
that the largest eigenvalue $\lambda^\ast$ is positive and simple, one has
to check whether there is an eigenvalue different from $\lambda^\ast$ with
real part $>\lambda^\ast/2$). This already explains on a formal level why,
for the ERW model, a phase transition occurs at $p=3/4$.

\section{Results and proofs for the standard ERW model}
\label{sec:results}
The paper of Janson~\cite{Ja} contains an exhaustive and very broad 
treatment of urn schemes and corresponding functional limit theorems. For
our purpose, it is most convenient to adapt the general results from 
there and to translate them into the setting of the ERW model, {\it
  via}~\eqref{eq:ERW-urn}. 

\subsection{The diffusive case ($0\leq p<3/4$)}
Our first convergence result deals with a distributional convergence of
processes, which holds in the Skorokhod space $D([0,\infty))$ of
right-continuous functions with left-hand limits. We simply recall that
distributional convergence in $D([0,\infty))$ to a process without
discontinuities at fixed times is stronger than finite-dimensional
distributional convergence, and point at~\cite{Bi} for more background.

\begin{theorem}
\label{thm:el-diffusive}
Let $0\leq p<3/4$. Then, for $n$ tending to
infinity, we have the distributional convergence in $D([0,\infty))$
$$
\left(\frac{S_{\lfloor tn\rfloor}}{\sqrt{n}}, t\geq 0\right)\Longrightarrow
  (W_t,t\geq 0),
$$
where $W=(W_t,t\geq 0)$ is a continuous $\R$-valued Gaussian process
specified by $W_0=0$, $\langle W_t\rangle=0$ for all $t\geq 0$, and
$$
\langle W_sW_t\rangle=\frac{s}{3-4p}\left(\frac{t}{s}\right)^{2p-1},\quad 0<s\leq t.
$$
\end{theorem}
We observe that when $p=1/2$, $W$ is a standard Brownian motion. Of course,
this we already know from Donsker's invariance principle, since in this
case, the ERW behaves as a simple symmetric (Bernoulli) random walk on
$\Z$, except possibly for the first step.

\begin{proof}
We  apply Theorem 3.31(i) of~\cite{Ja},
which shows that $$(n^{-1/2}(X_{\lfloor tn\rfloor}-tn\lambda_1v_1), t\geq
0)$$ converges in distribution towards a continuous $\R^2$-valued Gaussian
process $V=(V_t,t\geq 0)$ with $V_0=0$ and $\langle V_t\rangle = 0$ for all
$t\geq 0$. In our case, we have $\lambda_1=1$, and the covariance structure
of $V$ is closer specified in Remark $5.7$ of~\cite{Ja}. Display $(5.6)$
from that work shows that
$$
\langle V_s{V_t}'\rangle=s\Sigma_I\e^{\ln(t/s)A},\quad 0<s\leq t,
$$
with $\Sigma_I$ being a $2\times 2$-matrix defined under $(2.15)$
of~\cite{Ja}. An explicit calculation gives
$$\Sigma_I = \frac{1}{4(3-4p)}\begin{pmatrix}1&-1\\-1&1\end{pmatrix},$$
and the matrix exponential reads in our case
$$
\e^{\ln(t/s)A}=P\begin{pmatrix}\frac{t}{s}&0\\0&\left(\frac{t}{s}\right)^{2p-1}\end{pmatrix}P^{-1},\quad\hbox{with
} P=\frac{1}{2}\begin{pmatrix}1&1\\1&-1\end{pmatrix}.
$$
Together, we obtain for $0<s\leq t$,
$$
\langle V_s{V_t}'\rangle =
\frac{s}{4(3-4p)}\left(\frac{t}{s}\right)^{2p-1}\begin{pmatrix}1&-1\\-1&1\end{pmatrix}.
$$
By definition of $S_m$ and the continuous mappping theorem, we then deduce
that $(n^ {-1/2}S_{\lfloor tn\rfloor}, t\geq 0)$ converges in law in
$D([0,\infty))$ to a process $W=(W_t,t\geq 0)$ given by $W_t=V_t^1-V_t^2$
almost surely, where for $i=1,2$, $V^i$ denotes the $i$th component of
$V$. This proves our claim.
\end{proof}

Note that the covariance structure of the limit $W$ does not fit the
asserted effective diffusion coefficient in~\cite{ScTr}, cf. Display $(27)$
there. But the asymptotic behavior of the ERW mean square displacement
derived in~\cite{ScTr} (see Display~\eqref{eq:meansquare} above) is in
agreement with the second moment of $W$.

Moreover, we note that the initial steps of the ERW do not influence its
long-time behavior. Indeed, this can easily be derived from the fact that
the above urn admits the same Gaussian limit when starting from more
general configurations $\xi=(\xi^1,\xi^2)\in\N_0^ 2$ with
$\langle|\xi|^2\rangle <\infty$ and $\xi\neq (0,0)$. Specifying, for
example, to the deterministic initial configuration $\xi =(k_1,k_2)$ for
some $k_1,k_2\in\N$, the increment process $(X_n^1-X_n^2,n=1,2,\ldots)$ can
be seen as an ERW observed from time $k=k_1+k_2$ on when conditioned to be
at position $k_1-k_2$ at time $k$. Applying~\cite[Theorem 3.31(i)]{Ja} to
the urn when starting from configuration $\xi=(k_1,k_2)$, we deduce that
the first $k$ steps do not influence the limiting behavior.

\subsection{The critical case ($p=3/4$)}
In the borderline case $p=3/4$,  part (ii)
of~\cite[Theorem 3.31]{Ja} applies. 
\begin{theorem}
\label{thm:el-critical}
Let $p=3/4$. Then, for $n$ tending to
infinity, we have the distributional convergence in $D([0,\infty))$
$$
\left(\frac{S_{\lfloor n^t\rfloor}}{\sqrt{\ln n}\,n^{t/2}}, t\geq
  0\right)\Longrightarrow (B_t,t\geq 0),
$$
where $B=(B_t,t\geq 0)$ is a standard
one-dimensional Brownian motion. 
\end{theorem}
The function space $D([0,\infty))$ is defined as in the diffusive case
discussed above.

\begin{proof}
  According to Theorem 3.31(ii) of~\cite{Ja},
  $$((\ln n)^{-1/2}n^{-t/2}(X_{\lfloor n^t\rfloor}-n^t\lambda_1v_1), t\geq
  0)$$
  converges in law towards a continuous $\R^2$-valued Gaussian process
  $V=(V_t,t\geq 0)$ with $V_0=0$ and mean $\langle V_t\rangle=0$ for all
  $t\geq 0$. The covariance structure of $V$ is given by expression $(3.27)$
  of~\cite{Ja}, which simplifies in our case to
$$
\langle V_s{V_t}'\rangle=\frac{s}{4}\begin{pmatrix}1&-1\\-1&1\end{pmatrix},\quad
0<s\leq t.
$$
As above, the claim now follows from the continuous mapping theorem.
\end{proof}

Again, the asymptotics~\eqref{eq:meansquare} for the second moment of the
ERW obtained in~\cite{ScTr} match with the limit. With the same arguments
as in the diffusive case, one deduces moreover that the first steps of the
walker have no influence on the long-time behavior.

\subsection{The superdiffusive case ($3/4<p\leq 1$)}
In this regime, we can make use of Theorems 3.24 and 3.26
in~\cite{Ja}. 
\begin{theorem}
\label{thm:el-superdiffusive}
Set $\alpha=2p-1 \in(1/2,1]$. Then, for $n$ tending to infinity, we have
the almost sure convergence
$$
\left(\frac{S_{\lfloor tn\rfloor}}{n^\alpha}, t\geq 0\right)\longrightarrow
  (t^\alpha Y, t\geq 0),
$$
where $Y$ is some $\R$-valued random variable different from
zero. 
\end{theorem}
Below the proof of the theorem, we give some information on the limiting
variable $Y$.
\begin{proof}We note that in the notation of~\cite[Theorem 3.24]{Ja}, we
  have $\Lambda'_{\textup{III}}=\{2p-1\}$. We are therefore in the setting
  of the last part of the cited theorem and get
  that $$(n^{-\alpha}(X_{\lfloor tn\rfloor}-tn\lambda_1v_1), t\geq 0)$$
  converges almost surely to $(t^\alpha\hat{W}, t\geq 0)$, where
  $\hat{W}=(\hat{W}^1, \hat{W}^2)$ is some nonzero random vector lying in
  the eigenspace $E_{\lambda_2}$ of $A$, i.e.,
  $\hat{W}\in\{v\in \R^2: v= \lambda (1,-1)\textup{ for some
  }\lambda\in\R\backslash\{0\}\}$.
  Since $Y=\hat{W}^1-\hat{W}^2$ almost surely, the claim follows.
\end{proof}

In contrast to the regimes discussed in the two previous sections, the
distribution of $Y$ does depend on the law of the initial step of the
ERW. For example, in the degenerate case $p=1$, $Y$ has the same
distribution as $S_1=\xi^1-\xi^2$ (in fact,
$(S_{\lfloor tn\rfloor}=\lfloor tn\rfloor S_1)$ for all $t\geq 0$ with
probability one). In this regard, see also the remarks in~\cite{Ja} above
Theorem 3.9.

By looking at the skewness and kurtosis of the position of the walker for
large $n$, it was already observed in~\cite{SiCrScViTr} that the law of the
limit $Y$ cannot be Gaussian, even not when starting from the symmetric
initial condition $\P(\xi=(1,0))=\P(\xi=(0,1))=1/2$. See also~\cite{PaEs}
for a similar observation.

Moreover, we point at Theorem 3.26 of~\cite{Ja}, which can be used to
(recursively) calculate the moments of $Y$. Let us
for simplicity assume that $\xi=(1,0)$. Then, using additionally~\cite[Theorem
3.10]{Ja}, one finds for the first two moments
$$
\langle Y\rangle=\frac{1}{\Gamma(2p)},\quad \langle Y^2\rangle=\frac{1}{(4p-3)\Gamma(4p-2)},
$$
as we should have expected from Equations~\eqref{eq:mean}
and~\eqref{eq:meansquare}. For higher moments, see the remark
below Theorem 3.1 of~\cite{Ja}. We however mention that, even in the case of
an urn with deterministic replacement rules, there is in general no closed
form for the moments of the limiting variable. See~\cite{ChPoSa} and
further references therein for more on this.

\section{Extensions}
\label{sec:extensions}
It is the purpose of this section to exemplify that the approach {\it via}
P\'olya-type urns is robust and allows extensions and modifications of the
ERW model in various directions. We leave it to the reader to perform the
exact calculations and rather hint at the urn model one should consider.

\subsection{Higher dimensions}
Let us first explain how to obtain limit results for an ERW in higher
dimensions. In dimension $d\geq 1$, one should simply consider an urn with
$2d$ different colors. More specifically, in $d=2$, one might want to study
the urn $X_n=(X_n^1,X_n^2,X_n^3,X_n^4)$, $n\in\N$, with mean replacement matrix
$$
A_2=\begin{pmatrix}p&(1-p)/3&(1-p)/3&(1-p)/3\\
(1-p)/3&p&(1-p)/3&(1-p)/3\\
(1-p)/3&(1-p)/3&p&(1-p)/3\\
(1-p)/3&(1-p)/3&(1-p)/3&p
\end{pmatrix}.
$$
The corresponding nearest-neighbor ERW on $\Z^2$ is given by
$$
S_n=(X_n^1-X_n^2)e'_1 + (X_n^3-X_n^4)e'_2,
$$
with $e_1=(1,0)$ and $e_2=(0,1)$. Starting from $X_1=(1,0,0,0)$, say, this
means that the ERW first visits $(1,0)$. Then, at any later time $n\geq 2$,
the walker chooses a time $n'$ uniformly at random among the previous times
$1,\ldots,n-1$ and decides with probability $p$ to perform a step in the
same direction as at time $n'$, and with probability $(1-p)/3$ each to
perfom a step in one of the three other coordinate directions.

The expression for $S_n$ in the display above can again be analyzed with the results of
Janson~\cite{Ja}. In particular, since the eigenvalues of $A_2$ are given
by $\lambda_1=1$ and $\lambda_2=\lambda_3=\lambda_4=(4p-1)/3$, according to
the remarks before Section~\ref{sec:results}, a phase transition from
diffusive to superdiffusive behavior occurs at $p=5/8$.

\subsection{ERW with reinforced memory}
In a different direction, one might want to model an ERW which has a
reinforced memory, for example in the sense that the more often a
particular time from the past is remembered, the more likely it is to
remember this time again. From the point of view of neural networks, this
is certainly a reasonable and desirable assumption on the model. More
concretely, one might want to study a random walk with memory where the
remembered time $n'$ at the $n$th step is not chosen uniformly at random
among the previous times ${1,\ldots,n-1}$, but rather proportionally to a
weight distribution, with a weight that takes into account the number of
previous choices of $n'$.  In this regard, it is interesting to point at
the connection observed in~\cite{Kue} between the ERW model and so-called
random (uniform) recursive trees, which can naturally be used to model the
memory of the walker. The memory tree of an ERW with a reinforced memory
would correspond to a so-called preferential attachment tree; see,
e.g.,~\cite{DoMeSa}. In terms of a two-color urn, one might want to
consider a ``reinforced'' mean replacement matrix, for example
$$
B=\begin{pmatrix}a+p&1-p\\1-p&a+p\end{pmatrix},
$$
where $a\in\N_0$ is an additional parameter measuring the strength of the
reinforcement. Here, when a ball is drawn, one puts it back to the urn with
$a$ additional balls of the same color. In addition, one tosses a coin with
probability $p$ for heads and probability $1-p$ for tails. If a head shows up,
one adds another ball of the same color, whereas in case of tails, one puts
a ball of the opposite color into the urn. Note that the case $a=0$
corresponds to the uniform ERW model discussed above.

Again, this urn model fits into the general framework of urns treated
in~\cite{Ja}.  The eigenvalues of $B$ are given by $\lambda_1=a+1$ and
$\lambda_2=a+2p-1$. Hence, provided $a<3$, a phase transition for the urn
occurs at $p_a=(3-a)/4.$  

As above, let us now assume that the starting configuration of the urn is
given by a (possibly random) vector $\xi$ taking values in
$\{(1,0),\,(0,1)\}$. Regarding the corresponding random walk model
$S=(S_n,n\in\N_0)$ (we use the same notation as for the original ERW),
there is a little subtlety here: Most naturally, from time $1$ on, $S$
should not be defined as the difference $(X_n^1-X_n^2,n\in\N)$ of black and
red balls as before, but rather as the difference of black and red balls
which were put into the urn as a consequence of the coin tosses, plus the
initial difference $\xi^1-\xi^2$. In other words, one should not take into
account the $a$ additional balls of the same color which are put into the
urn at every draw for determining the position of the walker. In
particular, if $p=1/2$, except for the first step, $S$ behaves again like a
simple symmetric random walk (but note that $p_a<1/2$ if $a\geq 2$ !). If
$p\neq 1/2$, the behavior of the walk $S$ can be traced back to the
composition of the urn $((X_n^1,X_n^2),n\in\N)$. Namely, writing
$\Delta_n=S_{n+1}-S_n$ for the increment of the walker at time $n$, one
finds for its mean conditioned on $X_n$,
$$
\langle\Delta_n\rangle =(2p-1)\left(\frac{2X_n^1}{(a+1)n-a}-1\right).
$$
As to the urn, one can apply the results of~\cite{Ja} cited above to
obtain functional limit theorems, more precisely~\cite[Theorem 3.31(i)]{Ja}
in the case $p<p_a$,~\cite[Theorem 3.31(ii)]{Ja} in the case $p=p_a$,
and~\cite[Theorem 3.24]{Ja} in the case $p>p_a$. The usual diffusion approximation
now yields corresponding results for the walker $S$ when $p\neq 1/2$; namely, diffusive
behavior if $p<p_a$, marginally superdiffusive behavior if $p=p_a$ (with
the same rescaling as in Theorem~\ref{thm:el-critical}), and superdiffusive
behavior if $p>p_a$ (with the same rescaling as in
Theorem~\ref{thm:el-superdiffusive}).

\subsection{Modified ERW of Harbola, Kumar
  and Lindenberg~\cite{HaKuLi} }
Harbola, Kumar and Lindenberg~\cite{HaKuLi} proposed a modified ERW
representing a minimal one-parameter model of a random walk with memory,
which gives rise to all three possible types of behavior (superdiffusive,
diffusive and subdiffusive).  Again, $p\in [0,1]$ is a memory parameter
which is inherent to the model.

In contrast to the original ERW, the random walker moves only to the right,
but it may also stay still. More precisely, the modified ERW
$(S_n,n\in\N_0)$ starts at $S_0=0$, and then, at time $n\geq 1$, the
position of the walker is given by
$$
S_n = S_{n-1}+\sigma_n,
$$
with $\sigma_n$, $n\in\N$, being $\{0,1\}$-valued random variables with the
following law. Firstly, for concreteness, we assume that the first step goes
deterministically to the right, $\P(\sigma_1=1)=1$ (this is a slight
simplification compared with the model considered in~\cite{HaKuLi}).  At any
later time $n\geq 2$, we choose a number $n'$ uniformly at random among the
previous times ${1,\ldots,n-1}$. If $\sigma_{n'}=1$, i.e., the walker moved
to the right at time $n'$, we set
$$
\sigma_n= \left\{\begin{array}{l@{\quad\mbox{with probability  }}l}
      1 & p\\
      0&1-p
\end{array}\right..
$$
If $\sigma_{n'}=0$, i.e., the walker stood still at time $n'$, we set
$\sigma_n=0$, so that the walker does again not move at time $n$.

In the notation of Janson~\cite{Ja}, the mean replacement matrix of the
corresponding two-color urn (black balls for moving to the right, red balls
for standing still) is
$$
C=\begin{pmatrix}p&0\\1-p&1\end{pmatrix},
$$
where the first (second) column of $C$ is the expected change when a black
(red) ball is drawn. We stress that often in the literature (e.g., in~\cite{Ma})
rather the transpose $C'$ is considered as the mean replacement matrix.

In words, the dynamics of the urn process is described as follows: Starting
from some nontrivial initial condition at time $n=1$, we draw at time
$n=2,3,\ldots$, a ball uniformly at random, observe its color and put it
back to the urn. If we drew a black ball, we add with probability $p$
another black ball and with the complementary probability $1-p$ a red ball to the urn,
whereas if the observed color was red, we add deterministically another red
ball to the urn.

Note that if we start the urn model with one single black ball, the position $S_n$ of
the modified ERW at time $n$ is given by the number of black balls at time
$n$. 

The eigenvalues of the above matrix $C$ are $\lambda_1=1$ and
$\lambda_2=p$. Here, the results of~\cite{Ja} are not applicable, since
$\lambda_1$ does not belong to the dominating class: Indeed, when starting
the urn process from a single red ball, the dynamics adds only red balls to
the urn, and never a black ball. Such random triangular urn schemes were
however treated by Aguech~\cite{Ag}, generalizing the results of
Janson~\cite{Ja2} for triangular urns with deterministic replacement. In
particular,~\cite[Theorem 2(a)]{Ag} shows that the right rescaling for the
number of black balls at time $n$ is $n^p$ (there is no recentering), and
one has almost-sure convergence as $n$ tends to infinity to a nontrivial
(non-Gaussian) limit. This is in accordance with the results of Harbola,
Kumar, and Lindenberg~\cite{HaKuLi}, proving that in this random walk model,
subdiffusive (if $p<1/2$), diffusive (if $p=1/2$), and superdiffusive (if
$p>1/2$) behavior does occur.

A slightly more complicated model of a random walker moving to the left,
right, and staying put, which also exhibits all three types of behavior, was
presented by the same authors earlier in~\cite{KuHaLi}. There, one should
consider an urn with balls of three different colors: one corresponding to
a movement to the right, one corresponding to a movement to the left, and
one for staying at the same place.

\section{Conclusion}
In this paper we have explicitly determined the long-time behavior of the
one-dimensional ERW model introduced in 2004 by Sch\"utz and
Trimper~\cite{ScTr}. We used a simple connection to P\'olya-type urns and
relied on limit results for the latter that were already established
before. The ERW belongs to the class of models describing anomalous
diffusion and is one of the few models so far that turns out to be
explicitly solvable. As we exemplified in this paper, the ERW
model (and variants thereof) or, more generally, processes with
reinforcement can sometimes be reformulated in terms of urn models, which
have been studied for a long time in the mathematical literature and are
still objects of active research. In particular, results on urns often lead
to a deeper understanding of the corresponding random walk model.

\end{document}